\documentclass[%
superscriptaddress,
preprintnumbers,
nofootinbib,
twocolumn]{revtex4-1}

\def\aj{\ref@jnl{AJ}}                   
\def\actaa{\ref@jnl{Acta Astron.}}      
\def\araa{\ref@jnl{ARA\&A}}             
\def\apj{\ref@jnl{ApJ}}                 
\def\apjl{\ref@jnl{ApJ}}                
\def\apjs{\ref@jnl{ApJS}}               
\def\ao{\ref@jnl{Appl.~Opt.}}           
\def\apss{\ref@jnl{Ap\&SS}}             
\def\aap{\ref@jnl{A\&A}}                
\def\aapr{\ref@jnl{A\&A~Rev.}}          
\def\aaps{\ref@jnl{A\&AS}}              
\def\azh{\ref@jnl{AZh}}                 
\def\baas{\ref@jnl{BAAS}}               
\def\bac{\ref@jnl{Bull. astr. Inst. Czechosl.}}
\def\caa{\ref@jnl{Chinese Astron. Astrophys.}}
\def\cjaa{\ref@jnl{Chinese J. Astron. Astrophys.}}
\def\icarus{\ref@jnl{Icarus}}           
\def\jcap{\ref@jnl{J. Cosmology Astropart. Phys.}}
\def\jrasc{\ref@jnl{JRASC}}             
\def\memras{\ref@jnl{MmRAS}}            
\def\mnras{\ref@jnl{MNRAS}}             
\def\na{\ref@jnl{New A}}                
\def\nar{\ref@jnl{New A Rev.}}          
\def\pra{\ref@jnl{Phys.~Rev.~A}}        
\def\prb{\ref@jnl{Phys.~Rev.~B}}        
\def\prc{\ref@jnl{Phys.~Rev.~C}}        
\def\prd{\ref@jnl{Phys.~Rev.~D}}        
\def\pre{\ref@jnl{Phys.~Rev.~E}}        
\def\prl{\ref@jnl{Phys.~Rev.~Lett.}}    
\def\pasa{\ref@jnl{PASA}}               
\def\pasp{\ref@jnl{PASP}}               
\def\pasj{\ref@jnl{PASJ}}               
\def\rmxaa{\ref@jnl{Rev. Mexicana Astron. Astrofis.}}%
\def\qjras{\ref@jnl{QJRAS}}             
\def\skytel{\ref@jnl{S\&T}}             
\def\solphys{\ref@jnl{Sol.~Phys.}}      
\def\sovast{\ref@jnl{Soviet~Ast.}}      
\def\ssr{\ref@jnl{Space~Sci.~Rev.}}     
\def\zap{\ref@jnl{ZAp}}                 
\def\nat{\ref@jnl{Nature}}              
\def\iaucirc{\ref@jnl{IAU~Circ.}}       
\def\aplett{\ref@jnl{Astrophys.~Lett.}} 
\def\apspr{\ref@jnl{Astrophys.~Space~Phys.~Res.}}
\def\bain{\ref@jnl{Bull.~Astron.~Inst.~Netherlands}}
\def\fcp{\ref@jnl{Fund.~Cosmic~Phys.}}  
\def\gca{\ref@jnl{Geochim.~Cosmochim.~Acta}}   
\def\grl{\ref@jnl{Geophys.~Res.~Lett.}} 
\def\jcp{\ref@jnl{J.~Chem.~Phys.}}      
\def\jgr{\ref@jnl{J.~Geophys.~Res.}}    
\def\jqsrt{\ref@jnl{J.~Quant.~Spec.~Radiat.~Transf.}}
\def\memsai{\ref@jnl{Mem.~Soc.~Astron.~Italiana}}
\def\nphysa{\ref@jnl{Nucl.~Phys.~A}}   
\def\physrep{\ref@jnl{Phys.~Rep.}}   
\def\physscr{\ref@jnl{Phys.~Scr}}   
\def\planss{\ref@jnl{Planet.~Space~Sci.}}   
\def\procspie{\ref@jnl{Proc.~SPIE}}   

\usepackage{amssymb, amsmath, bm, dcolumn, epsf, graphicx, latexsym, slashed, simplewick}
\usepackage[utf8]{inputenc}

\usepackage{graphicx}
\usepackage{dcolumn}
\usepackage{bm}
\usepackage{xcolor}
\usepackage{bbold}
\usepackage{soul}
\usepackage[usestackEOL]{stackengine}
\usepackage{color}

\usepackage{amsmath}
\usepackage{amssymb}
\usepackage[normalem]{ulem}
\usepackage{float}

\usepackage{hyperref}

\usepackage{comment}

\begin{document}

\preprint{MIT-CTP/5725}
 
\title{\texttt{DiffLense}: A Conditional Diffusion Model for Super-Resolution \\of Gravitational Lensing Data}

\author{Pranath~Reddy}
\affiliation{University of Florida, Gainesville, FL 32611, USA}

\author{Michael~W.~Toomey}
\affiliation{Center for Theoretical Physics, Massachusetts Institute of Technology, Cambridge, MA 02139, USA}

\author{Hanna~Parul}
\affiliation{Department of Physics \& Astronomy, University of Alabama, Tuscaloosa, AL 35401, USA}

\author{Sergei~Gleyzer}
\affiliation{Department of Physics \& Astronomy, University of Alabama, Tuscaloosa, AL 35401, USA}

\date{\today}

\begin{abstract}
Gravitational lensing data is frequently collected at low resolution due to instrumental limitations and observing conditions. Machine learning-based super-resolution techniques offer a method to enhance the resolution of these images, enabling more precise measurements of lensing effects and a better understanding of the matter distribution in the lensing system. This enhancement can significantly improve our knowledge of the distribution of mass within the lensing galaxy and its environment, as well as the properties of the background source being lensed. Traditional super-resolution techniques typically learn a mapping function from lower-resolution to higher-resolution samples. However, these methods are often constrained by their dependence on optimizing a fixed distance function, which can result in the loss of intricate details crucial for astrophysical analysis. In this work, we introduce \texttt{DiffLense}, a novel super-resolution pipeline based on a conditional diffusion model specifically designed to enhance the resolution of gravitational lensing images obtained from the Hyper Suprime-Cam Subaru Strategic Program (HSC-SSP). Our approach adopts a generative model, leveraging the detailed structural information present in Hubble Space Telescope (HST) counterparts. The diffusion model, trained to generate HST data, is conditioned on HSC data pre-processed with denoising techniques and thresholding to significantly reduce noise and background interference. This process leads to a more distinct and less overlapping conditional distribution during the model's training phase. We demonstrate that \texttt{DiffLense} outperforms existing state-of-the-art single-image super-resolution techniques, particularly in retaining the fine details necessary for astrophysical analyses.
\end{abstract}

\maketitle

\section{Introduction}

Gravitational lensing, the bending of light from a distant source by a massive object between a source and the observer, is a powerful tool in astrophysics. Strong gravitational lensing in particular allows us to study the distribution of dark matter on subgalactic scales but also provides a magnified view of background sources which serves as a critical probe of the high redshift Universe. For detailed studies of background sources and the lens itself, high resolution and high quality data is imperative. However, the number of high-resolution  gravitational lensing data available is often limited in number, largely due to limitations in the capabilities of the observing instruments and adverse observing conditions. Thus, the generation of high-resolution data is imperative for future detailed studies of galaxies.

Despite these shortcomings, strong gravitational lensing has already shown significant potential in uncovering hints about the nature of dark matter through its substructures, evidenced by analyses of lensed quasars \cite{sub1,sub2,sub3}, observations from ALMA \cite{alma}, and extended lensing images \cite{veg,koop,veko}, among others. Indeed, various studies have explored anticipated signatures from $\Lambda$CDM and its extensions to derive information regarding the underlying dark matter distribution, e.g. \cite{Daylan:2017kfh,2010MNRAS.408.1969V,pcat,subs}.

Recently, there has been a surge in the use of machine learning to tackle questions in lensing \cite{Alexander:2019puy,Alexander:2021gxq,alexander_decoding_2020,Brehmer2019,Cranmer2020,Coogan2020,Montel2022,Hezaveh2017,Levasseur2017,Schuldt2021,Ostdiek2020, Lin2020, DiazRivero2020, Varma2020,Vattis:2020kaa,Ostdiek2022,Wagner-Carena2022}. Machine learning is well suited in this context as the analysis of even a single lens can be quite computationally taxing. Example applications of machine learning in this context include classification \cite{Alexander:2019puy, DiazRivero2020,Varma:2020kbq}, regression \cite{Brehmer:2019jyt,Levasseur2017}, segmentation analysis \cite{Ostdiek2022}, domain adaptation \cite{Alexander:2021gxq}, and anomaly detection \cite{alexander_decoding_2020}. So far, research has predominantly applied these techniques to simulations, primarily due to the limited availability of strong lensing data. This situation is expected to improve soon with the commissioning of the Vera C. Rubin Observatory and the launch of Euclid \cite{2019arXiv190205141V,2010MNRAS.405.2579O}. Most previous studies have relied on simulation data as a proxy for the absence of plentiful high quality lenses.  One possible work around to this issue is the implementation of super-resolution techniques applied to plentiful, lower quality data. 

Super-resolution techniques, particularly those based on machine learning, have shown promise in enhancing the quality of low-resolution astronomical images more generally \cite{superresolution1, superresolution2, superresolution3}.
Traditional methods typically involve learning a mapping from low-resolution (LR) to high-resolution (HR) images using a fixed distance function. This, however, can cause these methods to fail to capture intricate details essential for astrophysical analysis precisely because of this added rigidity.

To circumvent this, in this work, we introduce \texttt{DiffLense}, a novel super-resolution pipeline based on a conditional diffusion model. This model is specifically designed to enhance the resolution of gravitational lensing images obtained from the Hyper Suprime-Cam Subaru Strategic Program (HSC-SSP) \cite{hsc_ssp_pdr3}. \texttt{DiffLense} leverages the detailed structural information from high-resolution Hubble Space Telescope (HST) images to train the model. The diffusion model, trained on HST data, is conditioned on pre-processed HSC data to significantly reduce noise and background interference, ensuring a more distinct conditional distribution during training.

Specifically, our approach differs from traditional methods by adopting a generative approach, which better preserves the intricate details necessary for astrophysical analysis. We demonstrate that \texttt{DiffLense} outperforms existing state-of-the-art single-image super-resolution techniques, particularly in retaining fine details, thus providing a more accurate and detailed view of lensing morphology suitable for follow-up with traditional astrophysical analysis pipelines.

In Sec.~\ref{sec:Data}, we provide a comprehensive overview of the data sets utilized in this study. Sec.~\ref{sec:MT} details the models and methods employed in our analysis. We present the main findings in Sec.~\ref{sec:RES}, and conclude with a discussion and summary of our results in Sec.~\ref{sec:DNC}.

\section{Data} \label{sec:Data}

In this work we test our pipeline with two types of data sets for strong lensing. One we have constructed from real astrophysical observations and the second based on simulations. What is common between the two is that there is a set of low-resolution data of a source and a corresponding high-resolution image of the \textit{same} object.

\subsection{Real Lenses}

We have constructed a dataset containing images of strong galaxy-galaxy gravitational lenses observed with instruments with different resolution. We compiled a list of lens candidates from the literature \citep{Canameras_HSC, Canameras_panstarrs, Diehl, Garvin, Huang2020, Huang2021, Jacobs2019, Li, Pourrahmani, Rojas, Shu, Stein, Storfer, masterlens, Sugohi6, Sugohi7, Sugohi8} and crossmatched them with archival data. As a low resolution part, we utilized \textit{i}-band images from the third data release of Hyper Suprime-Cam Subaru Strategic Program (HSC-SSP), which has resolution of $0.168"$/pix. For high resolution counterparts we searched archival Hubble Space Telescope (HST) data available at MAST \footnote{\url{https://mast.stsci.edu/search/ui/}} and made cutouts from ACS/WFC images in F814W filter with $0.05"$/pix resolution. The final dataset contains 173 objects. 

\subsection{Simulated Lenses}

For our simulated data set we generate lenses with the \texttt{lenstronomy}\cite{Birrer:2018xgm} package. We model the dark matter halo with a spherical isothermal profile and produce lenses where the typical Einstein radius is $\sim 1.5''$. For modeling of the background galaxies we adopt a Sersic light profile and we further tune the apparent magnitude of the background galaxy such that the typical signal-to-noise ratio (SNR) of the lens arcs are consistent with real data, i.e. SNR $\sim$ 20 \cite{hst}. Furthermore, we construct these simulations to mimic the observing characteristics of HST by utilizing the default instrument and observing settings present in \texttt{lenstronomy}.

\section{Methodology} \label{sec:MT}
\subsection{Conditional Diffusion Model}

\begin{figure*}
    \centering
    \includegraphics[width=\linewidth]{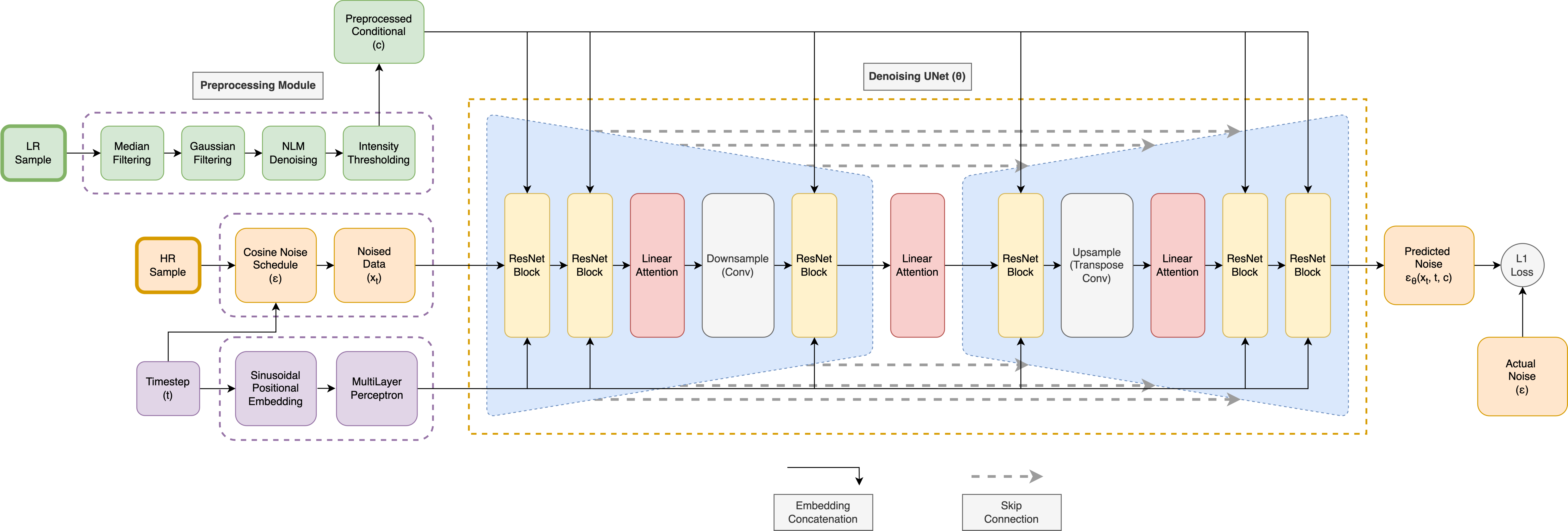}
    \caption{Overview of the \texttt{DiffLense} architecture.}
    \label{fig:block}
\end{figure*}

Diffusion models \cite{2020arXiv200611239H} represent a class of deep generative models that formulate a Markov chain to convert noise into a structured output. The core idea behind these models is inspired by non-equilibrium thermodynamics \cite{2015arXiv150303585S}, specifically the idea of reversing a diffusion process. The diffusion process or forward process here is modeled as a Markov chain that iteratively adds gaussian noise to the data, approximating the complex distribution of the image data with a series of simpler distributions, each corresponding to a different noise level. This process can be represented as,

\begin{equation}
    x_{t} = \sqrt{\alpha_{t}} x_{t-1} + \sqrt{1 - \alpha_{t}} \epsilon, \quad \epsilon \sim \mathcal{N}(0, I)
\end{equation}

where $x_{t}$ represents the data at time step $t$, $\alpha_{t}$ is the noise scale at step $t$, and $\epsilon$ is the Gaussian noise. The model learns to reverse this process, starting from a distribution of pure noise (high-entropy state) and progressively transitioning it into a distribution that closely resembles the target structured data (low-entropy state). This is achieved by training a neural network to predict the reverse diffusion steps, effectively learning to de-noise the data at each step to eventually yield a coherent image or pattern. The reverse diffusion can be formulated as,

\begin{equation}
    x_{t-1} = \frac{1}{\sqrt{\alpha_{t}}} \left( x_{t} - \frac{1 - \alpha_{t}}{\sqrt{1 - \alpha_{t}}} \epsilon_{\theta}(x_{t}, t) \right)
\end{equation}

where $x_{t-1}$ is the denoised data at time step $t-1$, $x_{t}$ is the noised data at time step $t$, and $\epsilon_{\theta}(x_{t}, t)$ is the predicted noise at time step $t$, parameterized by the neural network with parameters $\theta$. Conditional diffusion models extend the generative capabilities of diffusion models by conditioning the generative process on additional information $c$.  

\begin{equation}
    x_{t-1} = \frac{1}{\sqrt{\alpha_{t}}} \left( x_{t} - \frac{1 - \alpha_{t}}{\sqrt{1 - \alpha_{t}}} \epsilon_{\theta}(x_{t}, t, c) \right)
\end{equation}

This conditioning can be based on various types of auxiliary inputs, such as class labels \cite{Le2023}, images \cite{2021arXiv211105826S}, text descriptions \cite{2022arXiv220406125R, 2022arXiv220511487S}, or, as in our case, low-resolution images. By conditioning on these inputs, the model can generate data that is relevant to the given context. In our methodology, the diffusion model is conditioned on low-resolution images from the Hubble Space Telescope, which provides the context for the high-resolution image it needs to generate. This approach ensures that the generated high-resolution images maintain the astrophysical characteristics of the original low-resolution images. 

\subsection{Implementation}

The model architecture is based on a U-Net structure \cite{2015arXiv150504597R}, a  convolutional neural network known for its effectiveness in image-to-image translation tasks. The U-Net model consists of a series of downsampling and upsampling layers interconnected with residual connections \cite{2015arXiv151203385H}. These connections are crucial in preserving and propagating detailed spatial information throughout the network. The downsampling layers capture contextual information from the input image, while the upsampling layers incrementally increase the resolution, refining and adding details at each step. The randomly sampled time step is also passed to the model to help the model learn the noise distribution of discrete time steps. This U-Net architecture is inspired by the model used in \cite{10.1093/mnras/stac130}, given the success of this architecture in generating realistic galaxy images.

In our implementation, instead of the traditional approach of passing in a latent space vector \cite{2021arXiv211210752R} or concatenating the conditional image with the noised input of the U-Net \cite{9887996} or incorporating an encoding of the conditional image along with the time step encoding, the model directly concatenates the conditional image at every ResNet block within the U-Net. By concatenating the conditional image at every ResNet block, we ensure that the low-resolution information is actively utilized throughout the model. This leads to a more integrated and consistent use of conditional data, improving the feature mapping from low to high-resolution samples. This method allows for continuous contextual guidance during the denoising process. In astrophysical imaging, where subtle features and fine details are crucial, having continuous low-resolution context helps in preserving these details more effectively. The system architecture of \texttt{DiffLense} is presented in Figure~\ref{fig:block}.

\begin{figure*}[!t]
    \centering
    \includegraphics[width=\linewidth]{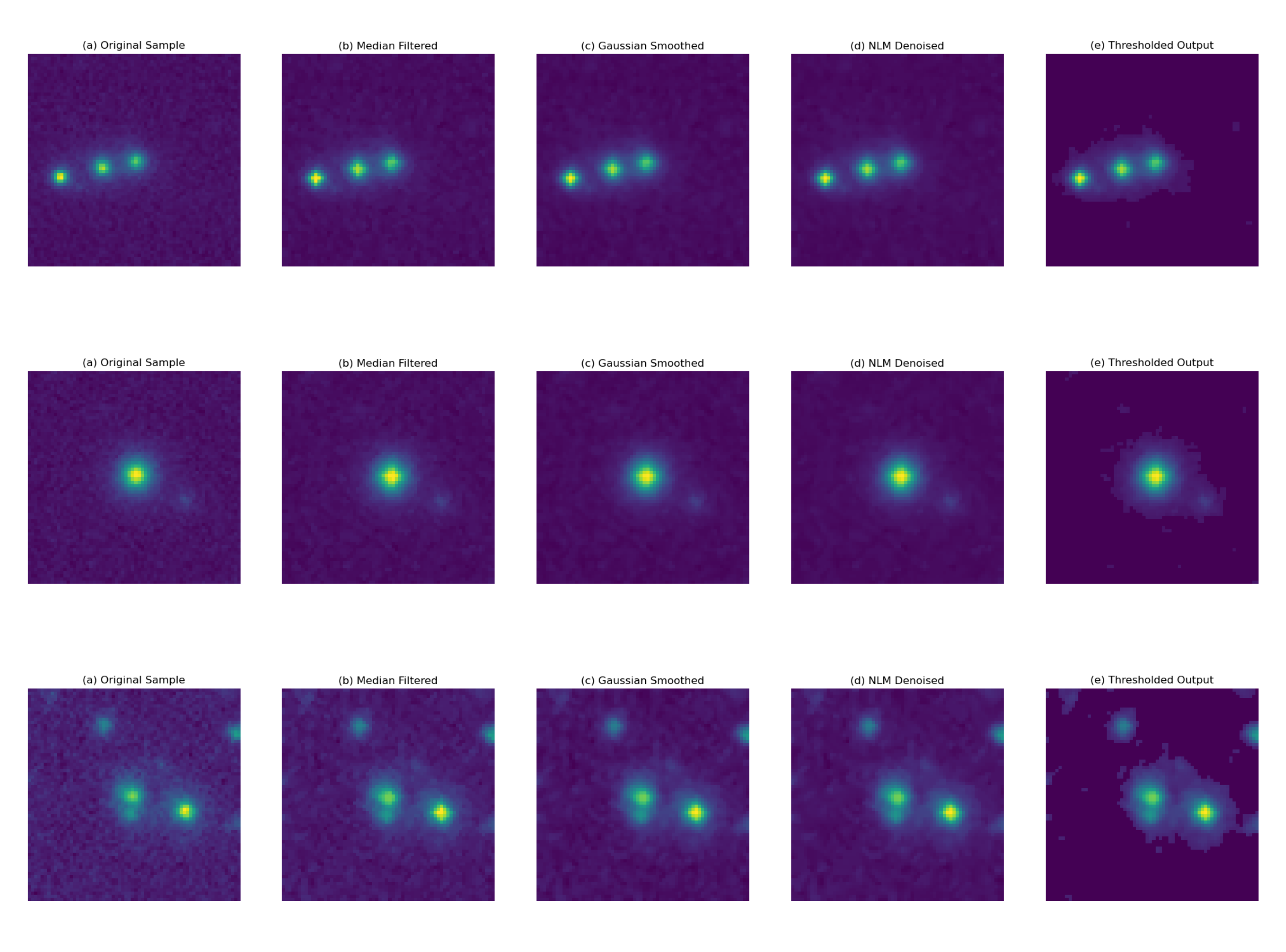}
    \caption{Pre-processing examples of conditional inputs. From left to right: HSC input sample, Median Filtering output, Gaussian Filtering output, Denoising output, and output of Thresholding.}
    \label{fig:preproc}
\end{figure*}

\subsection{Pre-processing of Conditional Inputs}

A crucial step in the pipeline is the pre-processing of conditional inputs, specifically the low-resolution HSC images. This pre-processing stage is vital for reducing noise and background interference, which significantly contributes to a more distinct and less overlapping conditional distribution during the model's training phase providing a clearer context for each step of the reverse diffusion process.

The pre-processing pipeline for HSC images involves several key steps. The first step involves applying a median filter to the images, which helps in removing salt-and-pepper noise. This is followed by a Gaussian filter with a sigma value of 0.5, which smoothens the image by blurring out small irregularities and noise. The combination of these two filters effectively reduces the random noise in the images without significantly altering the underlying distribution of the data. 

After initial smoothing, we apply a denoising step using Non-Local Means (NLM) Denoising \cite{ipol.2011.bcm_nlm}. Non-Local Means Denoising works by comparing all patches in the image and averaging similar ones. Post-denoising, we normalize the images using min-max normalization, which is important for standardizing the data. Lastly, to reduce the background interference, a thresholding technique is applied. The thresholding value is set based on the mean and standard deviation of the images and pixels with intensity values below this threshold are set to zero, effectively suppressing the background noise. Examples demonstrating the application of each pre-processing step are presented in Figure \ref{fig:preproc}. 

\subsection{Experimental Setup}

The U-Net architecture as shown in Figure \ref{fig:block} consists of several components such as the upsampling and downsampling layers for processing feature maps at different scales, residual blocks and skip connections for enabling the flow of gradients through deeper layers without attenuation, sinusoidal positional embeddings for incorporating temporal information into the model that is crucial for the diffusion process, and linear attention mechanism to efficiently compute attention over images. The activation function used is the Mish Activation \cite{2019arXiv190808681M}, a smooth, non-monotonic function that helps maintain the flow of gradients and prevent vanishing gradient issues. 

To control the progression and noise levels during the diffusion process, we use the cosine noise scheduler \cite{2021arXiv210209672N}, a schedule that uses the cosine function to vary the beta values (noise levels) over 1000 timesteps. This schedule allows for a smoother transition of noise levels compared to a linear noise scheduler, potentially leading to more stable training and cleaner sampling of outputs.

The dataset includes $2880$ HR and LR lensing image pairs. HR images are $128 \times 128$ pixels, while pre-processed conditional LR images are $64 \times 64$ pixels. Both the HR and LR samples are normalized to the range $[-1,1]$ and passed to a data loader with the batch size set to $10$. The model is trained for $2000$ epochs using the the Adam optimizer with a learning rate of $2\times10^{-5}$ for minimizing the L1 Loss function that calculates the difference between the actual and the predicted noise in the input. The model is implemented using the \texttt{PyTorch} package \cite{2019arXiv191201703P} and trained on two NVIDIA Tesla A100 GPUs.

\section{Results} \label{sec:RES}

\subsection{Evaluation of Existing Super-Resolution Models} \label{sec:Baseline}

In the initial phases of our research, we have conducted an in-depth evaluation of four established single-image super-resolution models. We train the models to map HST data to its HSC counterparts. This phase is important in establishing a baseline against which the performance of our proposed method, \texttt{DiffLense}, could be measured. These models – Residual Dense Network (RDN), Residual Channel Attention Network (RCAN), Swin Image Restoration Transformer (SwinIR), and Hybrid Attention Transformer (HAT) – chosen for this analysis have a unique set of capabilities that are advantageous for addressing the complex challenges of astrophysical imaging.

The Residual Dense Network (RDN) \cite{8578360} model effectively captures local features through densely connected convolutional layers. The key idea is to use residual learning at two levels: local (within the dense blocks) and global (between the input and output of the entire network). This helps in preserving the original image details while enhancing resolution. The Residual Channel Attention Network (RCAN) \cite{2018arXiv180702758Z} improves upon traditional convolutional networks by introducing channel attention mechanisms within each residual block. This allows the network to focus on more informative channel features by adaptively rescaling channel-wise features. It is effective in handling real-world images where certain features may be more important than others for reconstruction.

SwinIR \cite{2021arXiv210810257L} leverages the Swin Transformer \cite{2021arXiv210314030L}, which uses shifted windows to limit self-attention computation to non-overlapping local windows while also allowing for cross-window connections. This results in an efficient and scalable method that performs well on image restoration tasks, including super-resolution. Finally, HAT \cite{2022arXiv220504437C} combines the transformer architecture \cite{2017arXiv170603762V}  with hybrid attention mechanisms, allowing it to capture both local and global dependencies effectively. It integrates the local feature extraction capabilities of CNNs with the long-range dependency modeling of Transformers. This hybrid design allows the model to focus on both the low-level and high-level features, enhancing the model's ability to reconstruct high-resolution details from low-resolution images.

We have used the Adam optimizer \cite{2014arXiv1412.6980K} for training the models with Mean Absolute Error (MAE) as our loss function,

\begin{equation}
    \text{MAE} = \frac{1}{n} \sum_{i=1}^{n} |y_i - \hat{y}_i|
\end{equation}

\begin{figure*}[!t]
    \centering
    \includegraphics[width=\linewidth]{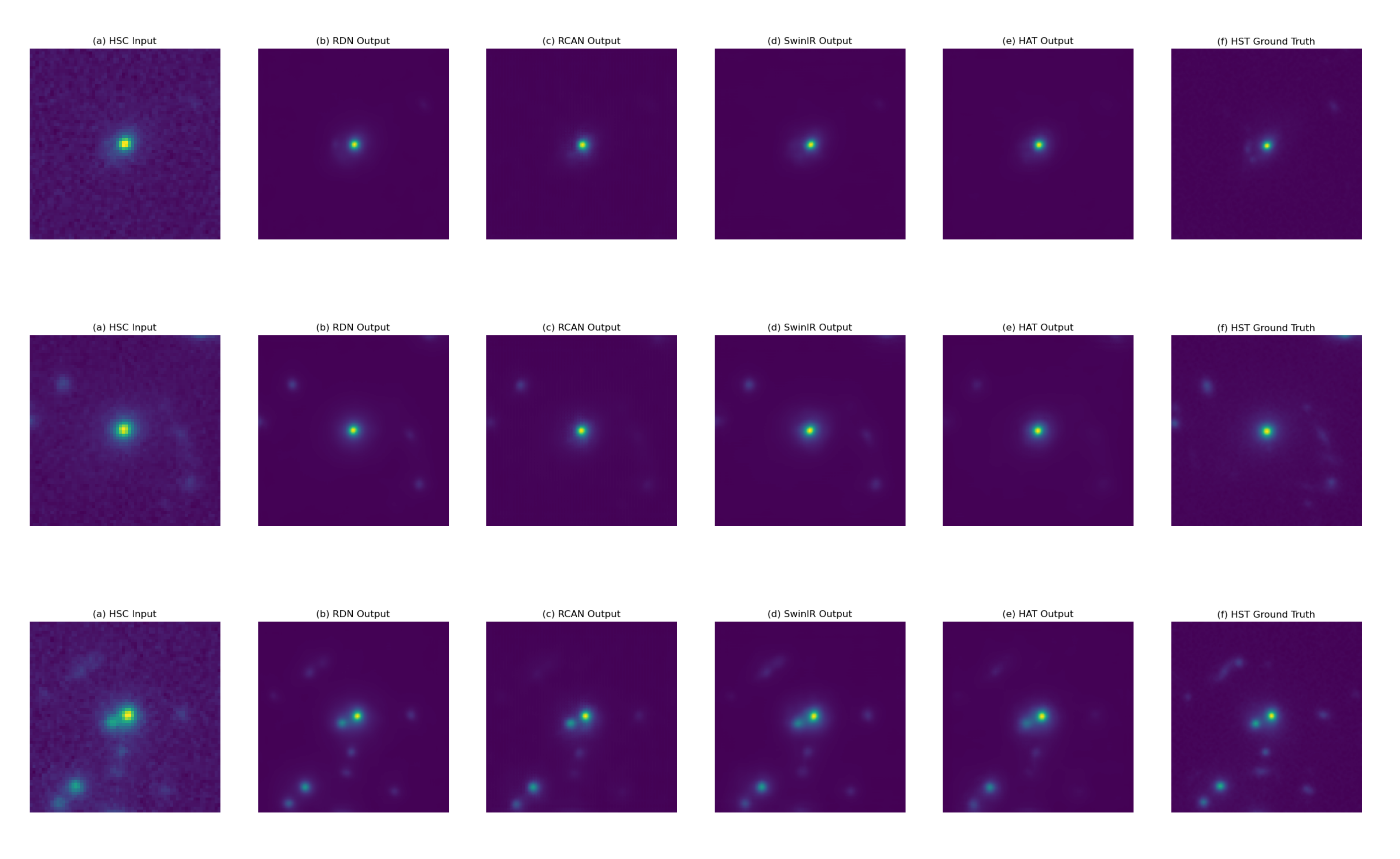}
    \caption{Super-resolution examples from preliminary analysis. From left to right: HSC input sample, RDN output generation, RCAN output generation, SwinIR output generation, HAT output generation, and the HST ground truth sample.}
    \label{fig:prelim}
\end{figure*}

where $y_i$ represents the actual observed values and $\hat{y}_i$ represents the predicted values. We set an initial learning rate of $2\time10^{-3}$, with no weight decay, and beta parameters for the optimizer at $0.9$ and $0.99$. In determining the number of training epochs, we have followed an adaptive approach, setting the epoch count based on the convergence speed of each model. This allows us to set the training duration depending on the specific learning characteristics and optimization requirements of each model. A cyclic learning rate scheduler \cite{2015arXiv150601186S} is employed to adjust the learning rate throughout the training process, optimizing the convergence speed and stability. During training, each epoch consists of a forward pass of the model with the LR input data, followed by a loss calculation using the HR target data. This loss is then backpropagated to update the model parameters.

The post-training evaluation phase involved assessing each models' performance using a comprehensive set of quantitative metrics, including Mean Squared Error (MSE), Mean Absolute Error (MAE), Structural Similarity Index Measure (SSIM) \cite{1284395}, and Peak Signal-to-Noise Ratio (PSNR). The PSNR metric is a logarithmic measure that quantifies the ratio of the maximum possible intensity of an image signal to the noise. A higher PSNR typically indicates lower distortion and is thus used as a standard criterion for evaluating the fidelity of image reconstruction algorithms. SSIM is a function of the mean intensity, contrast variance, and covariance, offering a more perceptually relevant assessment of image quality than traditional error summation methods like MSE, which PSNR is based upon.

These metrics offer a holistic view of the models' capabilities, allowing us to gauge not only the accuracy but also the perceptual integrity and detail preservation in the super-resolved images. In addition to the quantitative metrics, we have also conducted a qualitative analysis by performing visual inspections of the model outputs. It involves a detailed examination of the models' outputs, comparing them visually against the original high-resolution images. The quantitative results are summarized in Table \ref{table:sr_models}, and the output visualizations are presented in Figure \ref{fig:prelim}.

\begin{table}[h]
\centering
\caption{Comparative Performance of Super-Resolution Models}
\begin{tabular}{lcccc}
\hline
Model & MAE & MSE & SSIM & PSNR \\ \hline
RDN & 0.03648 & 0.01585 & 0.86701 & 32.94286 \\
RCAN & 0.03671 & 0.01577 & 0.86768 & 33.80036 \\
SwinIR & 0.03653 & 0.01576 & 0.86866 & 34.63678 \\
HAT & 0.03677 & 0.01584 & 0.86565 & 34.01312 \\ \hline
\end{tabular}
\label{table:sr_models}
\end{table}

The results from this analysis present a comprehensive overview of each model's performance. SwinIR emerged slightly ahead in terms of SSIM and PSNR, indicating a superior reconstruction of complex image structures and textures. Furthermore, all models succeeded in effectively eliminating background noise. However, despite these encouraging outcomes, it is evident from the visualizations that the models have limitations in fully capturing the complex astrophysical features essential for a comprehensive analysis of the gravitational lenses. These shortcomings are particularly noticeable in the reconstruction of the lensed galaxies, an area where precision and detail fidelity are critical. While effective in enhancing the overall image resolution, the models struggled with preserving the fine details. Balancing the reconstruction of the structure with noise reduction emerged as a common issue across the models, which is a result of the models focusing solely on optimizing a distance-based objective function such as MAE. This evaluation highlights the necessity for an advanced approach capable of not only upscaling images but also adequately preserving the astrophysical details. This realization led to the development of \texttt{DiffLense}. Our approach aims to fill the gaps identified in these preliminary models, leveraging a generative approach to achieve better clarity and detail in the super-resolved astrophysical images.

\subsection{Performance of \texttt{DiffLense}}

\begin{figure*}[!t]
    \centering
    \includegraphics[width=\linewidth]{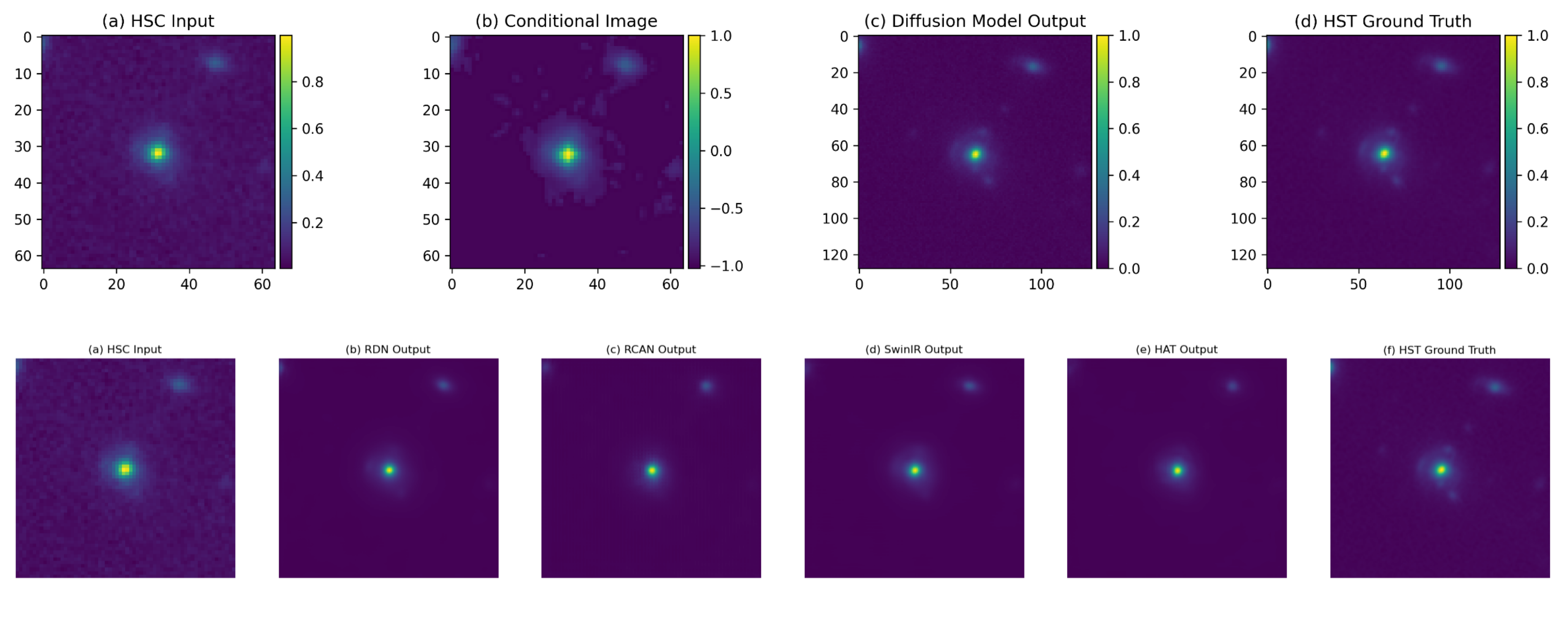}
    \caption{\texttt{DiffLense} Output Example. Top: HSC input, the pre-processed conditional, and the diffusion model output. Bottom: Outputs of the same sample from the baseline models used in preliminary analysis.}
    \label{fig:diff_sample}
\end{figure*}

Having established the limitations of current super-resolution methods, we now turn our attention to the performance of \texttt{DiffLense}. In this section, we analyze the performance of our method and also compare the results with the four baseline models discussed in the above subsection. The baseline models are proficient at denoising and smoothing the HSC input images. However, they exhibit a tendency to produce outputs that, despite being cleaner, lack the structure and detail that are vital for a detailed interpretation of the lenses. These models fail to maintain the balance between noise reduction and the preservation of fine details, resulting in images that are overly smooth. In comparison, the outputs generated by \texttt{DiffLense} are not only less noisy but also showcase better detail. As seen in an example presented in Figure \ref{fig:diff_sample}, \texttt{DiffLense} does a much better job at reconstructing the lensed galaxies around the central lens. 

We extend this analysis through quantitative analysis, using metrics such as SSIM and PSNR. Our method achieved an SSIM of $0.83937$ and a PSNR of $35.06834$. While the SSIM value does not correlate well with the observations, the PSNR indicates a significant improvement in image quality. The SSIM measures the similarity between two images, which includes attributes such as texture, contrast, and structure. The reduced SSIM score could potentially be attributed to the misalignment in the intensity range between the super-resolved images and the ground truth. This discrepancy could be due to the generative nature of the diffusion model. Unlike baseline models that directly calculate loss between predicted and actual images leading to better intensity alignment, the diffusion model clamps the intensities to a fixed range of $[-1,1]$ and normalizes them, potentially causing some slight misalignment. Additionally, the model's reverse process, which iteratively predicts and subtracts the noise based on the outputs from prior time steps, could lead to the compounding of errors. Inaccuracies in early steps may escalate, increasing the numerical error observed in later outputs. Although the SSIM score does not correlate well with the perceived visual quality, the PSNR score is significantly improved, indicating superior reconstruction quality and reduction of noise. 

\begin{figure*}[!t]
    \centering
    \includegraphics[width=\linewidth]{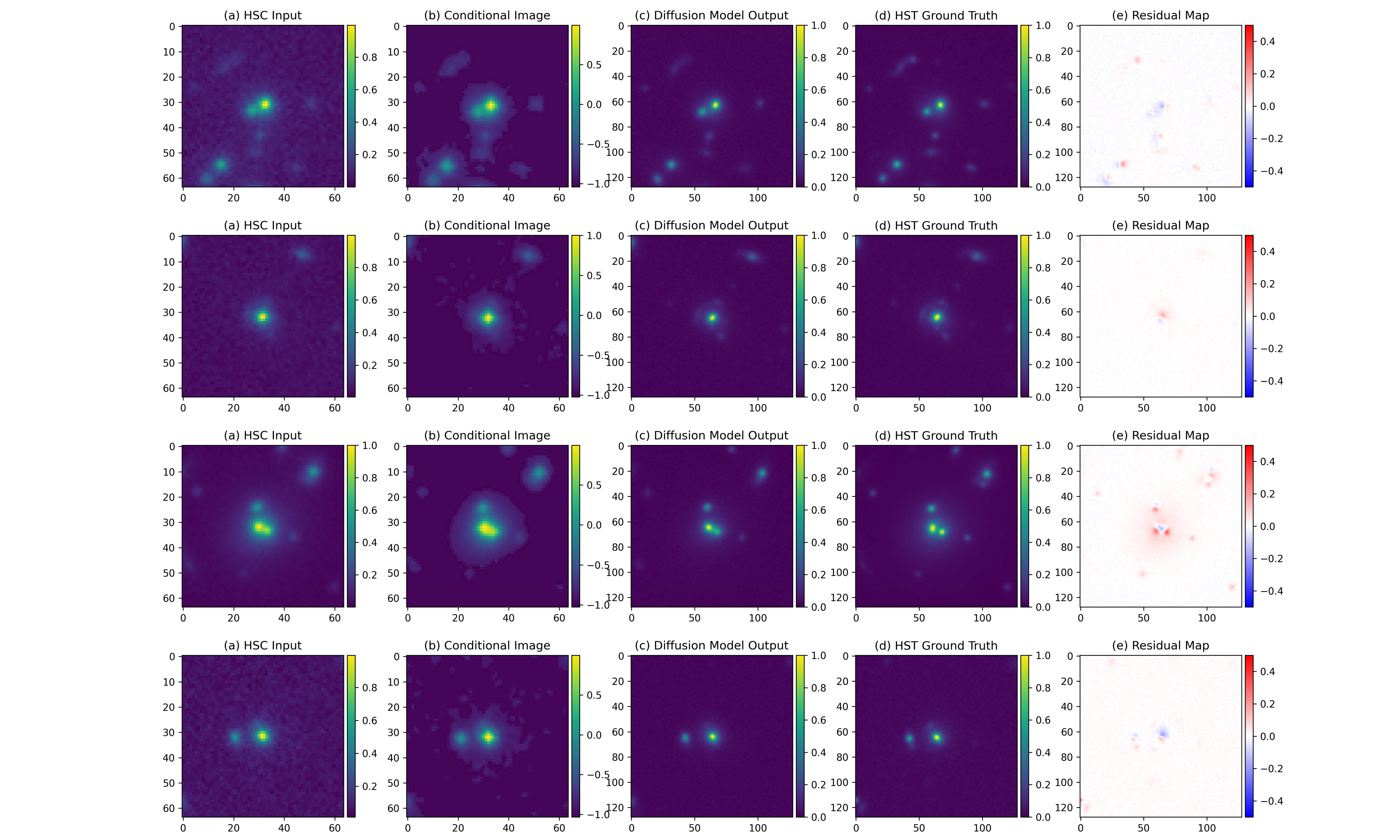}
    \caption{\texttt{DiffLense} Output Examples w/ Residual Maps. From left to right: HSC input sample, conditional image, \texttt{DiffLense} output, ground truth, and the residual map.}
    \label{fig:diff_res}
\end{figure*}

We further extend this analysis by examining the residual maps, as shown in Figure \ref{fig:diff_res}. These maps are calculated using the differences in intensity between the actual and predicted images normalized by the former. These maps are a direct measure of the model's performance in replicating the ground truth. While some of the errors could be attributed to the misalignment of the intensities as discussed earlier or to the residual noise from the reverse process, we see a slightly positive difference in the background of the images, which indicates a reduction in the background noise. However, the model seems to be overestimating and underestimating the spread of the light. We notice consistent dark and bright spots that are symmetrically distributed around the central objects, hinting that the model's error may be consistent in how it handles the light profile around these objects. This highlights a potential area for improvement and a requirement for a deeper study of the model's performance.

\subsection{Performance on Simulation Dataset}

Following the evaluation of the super-resolution models using real astronomical datasets, we extend our analysis to the simulated dataset to evaluate the model performance on a larger training sample size. This dataset includes HR images that are generated following the simulation process detailed in Section \ref{sec:Data}, and the LR counterparts are produced by a two-stage degradation process. We first apply Gaussian blurring to the high-resolution images. This blurring process, achieved by convolving the images with a Gaussian kernel whose standard deviation varies randomly within the range $(0.5, 2.5)$, simulates the optical and atmospheric distortions typical of ground-based astronomical observations. Following this, Gaussian noise, whose standard deviation varies randomly within the range $(0.01, 0.1)$, was added to introduce the typical sensor and environmental noise observed in real imaging scenarios. We present the quantitative performance of \texttt{DiffLense} alongside the established baseline models in Table \ref{table:sr_models_sim}.

\begin{table}[h]
\centering
\caption{Comparative Performance of Baseline Models}
\begin{tabular}{lcccc}
\hline
Model & MAE & MSE & SSIM & PSNR \\ \hline
RDN & 0.00230 & 0.00013 & 0.9775 & 40.304 \\
RCAN & 0.00229 & 0.00012 & 0.97908 & 40.272\\
SwinIR & 0.00204 & 0.00011 & 0.98059 & 40.391\\
HAT & 0.00185 & 0.00010 & 0.98352 & 40.709\\
\texttt{DiffLense} & 0.00558 & 0.00017 & 0.98422 & 39.189\\
\end{tabular}
\label{table:sr_models_sim}
\end{table}

The results showcase a distinctive performance profile for the \texttt{DiffLense} model. Notably, while it exhibits higher L1 Loss and MSE values, which can be attributed to residual noise inherent in the reverse diffusion process, it achieves the highest SSIM among all the models suggesting a superior preservation of structural information within the image.

\begin{figure*}[!t]
     \centering
     \includegraphics[width=\linewidth]{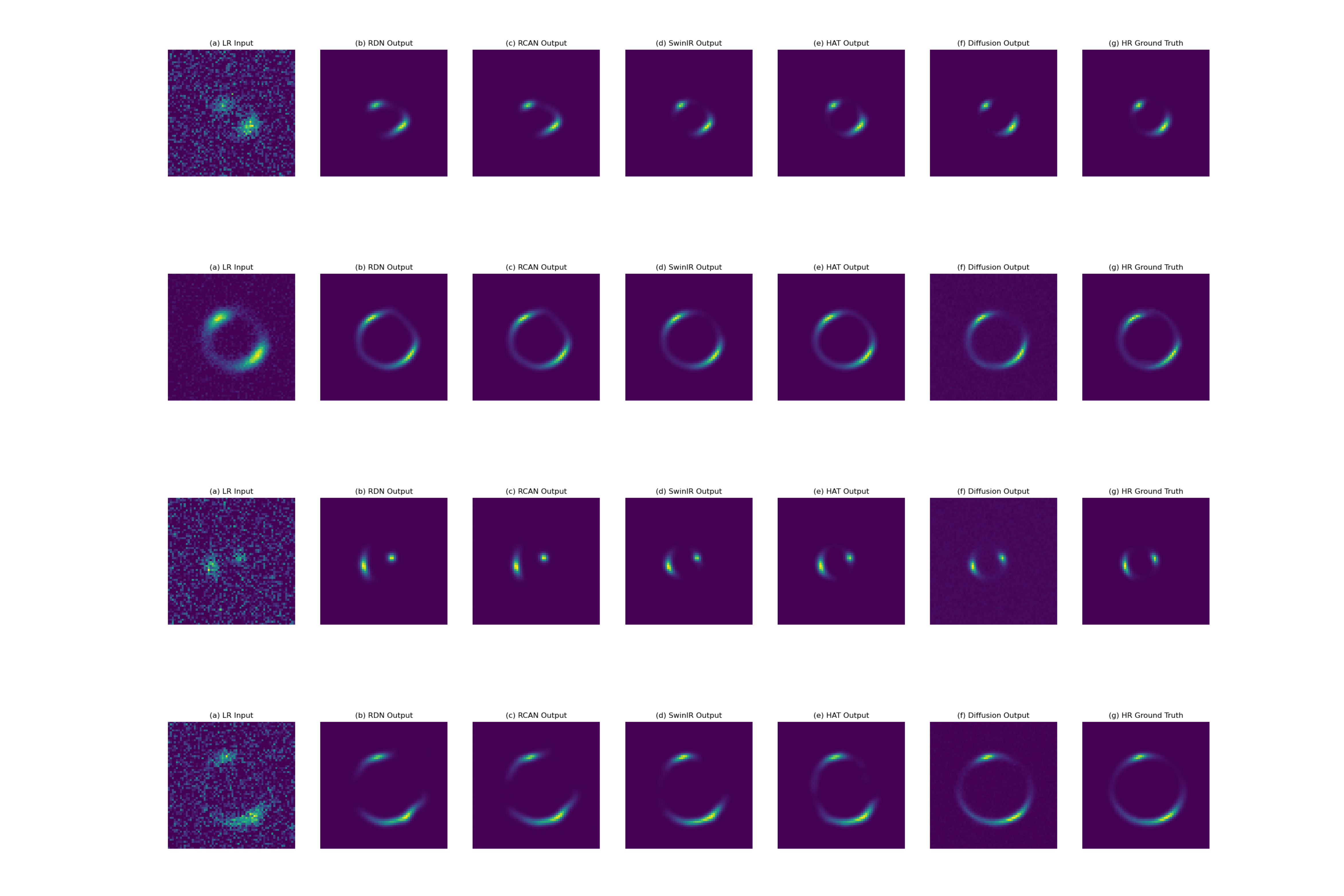}
     \caption{Super-resolution examples from the simulation dataset. From left to right: LR input sample, RDN output generation, RCAN output generation, SwinIR output generation, HAT output generation, \texttt{DiffLense} output generation, and the HR ground truth sample.}
     \label{fig:results_sim}
\end{figure*}

Outputs presented in Figure \ref{fig:results_sim} offer a visual comparison of the model's performance. These results highlight the diffusion model's proficiency in preserving the structural integrity of the lensed galaxies. In comparison, the simulation results show that the baseline models have slightly distorted lens structures even if they have managed to remove part of the smoothness seen in case of the real lenses.

\section{Discussion \& Conclusion} \label{sec:DNC}

The application of super-resolution techniques to gravitational lensing images represents a potentially significant advancement in astrophysical imaging. In this study we introduced \texttt{DiffLense}, a novel super-resolution pipeline based on a conditional diffusion model designed to enhance the resolution of gravitational lensing images. Our method significantly improves the resolution of images obtained from HSC-SSP by leveraging the detailed structural information from high-resolution HST images. We also demonstrate the generality of the approach in the controlled domain of simulation.

Our results demonstrate that \texttt{DiffLense} outperforms existing state-of-the-art single-image super-resolution techniques, particularly in preserving the fine details necessary for astrophysical analysis. Traditional methods often fall short in capturing intricate details due to their reliance on optimizing a fixed distance function. In contrast, \texttt{DiffLense} adopts a generative approach that better preserves these details, leading to more accurate and detailed super-resolved images. 

\begin{itemize}
    \item \texttt{DiffLense} effectively restores the intricate structures and fine details in gravitational lensing images, which are critical for accurate astrophysical analysis. This capability is demonstrated through quantitative metrics and explicit inspections, where \texttt{DiffLense} consistently produced higher quality images compared to baseline models.
  
    \item The pre-processing pipeline for conditional inputs significantly reduces noise and background interference, resulting in a clearer and more distinct conditional distribution during the model's training phase. This leads to a more accurate final output.
  
    \item By using a conditional diffusion model, \texttt{DiffLense} benefits from the generative process, which allows for a more nuanced and detailed reconstruction of high-resolution images from low-resolution inputs. This approach outperforms traditional methods that rely on direct optimization of distance metrics.

    \item The refinement of diffusion models like \texttt{DiffLense} opens the possibility for expanding the prevalence of high resolution data.
\end{itemize}

The development of \texttt{DiffLense} marks a step forward in the application of machine learning techniques to astrophysical imaging. By leveraging the strengths of conditional diffusion models, \texttt{DiffLense} provides a powerful tool for enhancing the resolution of gravitational lensing images, which in the future could enable more accurate scientific analyses of gravitational lenses. This work not only demonstrates the potential of generative models in this domain but also paves the way for future innovations in applying super-resolution methods to astrophysical imaging for other applications.

\section*{Acknowledgements}
We acknowledge useful conversations with Stephon Alexander. P.R. was a participant in the Google Summer of Code 2023 program.  S.G. was supported in part by U.S. National Science Foundation award No. 2108645. Portions of this work were conducted in MIT’s Center for Theoretical Physics and partially supported by the U.S. Department of Energy under grant Contract Number DE-SC0012567. M.W.T is supported by the Simons Foundation (Grant Number 929255).

\bibliography{bib.bib}


\end{document}